\documentstyle[prb,aps,epsf,epsfig]{revtex}
\begin{document}
\draft
\twocolumn

\title{
Manifestation of quantum chaos on ordered
structures by scattering techniques:
application to Low-Energy Electron Diffraction 
}

\author{
P. L. de Andres 
and 
J. A. Verg\'es
}

\address{
Instituto de Ciencia de Materiales de Madrid,
Consejo Superior de Investigaciones Cient\'{\i}ficas,
Cantoblanco, E-28049 Madrid, Spain} 

\date{\today}

\maketitle

\begin{abstract}
\baselineskip=2.5ex
We analyze statistical probability distributions
of intensities collected by
diffraction techniques like
Low-Energy Electron Diffraction.
A simple theoretical model based in
hard-sphere potentials and LEED formalism is investigated
for different values of relevant parameters: energy,
angle of incidence, muffin-tin
potential radius, maximum spherical
component $l_{max}$, 
number of stacked layers, 
and full multiple-scattering
or kinematic model. 
Given a complex enough system
(e.g., including multiple scattering
by at least two Bravais lattices),
the computed probability distributions
agree rather well with a $\chi^{2}_{2}$ one,
characteristic of the Gaussian Unitary Ensemble
universality class associated to quantum chaos.
A hypothesis on the possible
impact of the chaoticity of wavefunctions
on correlation factors is tested against the
behaviour of 
the Pendry R-factor and the Root Mean Squared
Deviation factor.
\end{abstract}


\narrowtext

\section{introduction}


There is much interest to understand the role of
chaos on quantum systems. 
The important consequences of chaos on classical systems
yield a clear motivation to extend the study of chaos
to the quantum physics world.
It is difficult, however, to explore quantum
chaos by connecting the quantum and classical formulation
through some special limits\cite{berrylim}.
Therefore, a group of pioneering investigators adopted
long time ago a point of view well independent of any
classical or semiclassical approach to chaos\cite{wigner-dyson}. 
In this approach,
the object of interest is a Hamiltonian composed of 
random numbers (RMT)\cite{mehta,brody}, 
conjectured by Wigner, Dyson and others to be a relevant
prototype for quantum chaotic behaviour. A further
conjecture by Porter and Thomas\cite{porter} established the probability
distribution to be expected for intensities related to a typical chaotic
wavefunction: $\chi^{2}_{\nu}$. 
This is a function that gives the probability distribution of
intensities $I/<I>$, over the spatial support of the wavefunction
at a given energy
($<I>$ is the corresponding mean value).
Later on, Dyson demonstrated that the parameter
$\nu$ can only take three different values
(i.e., $1$, $2$, and
$4$, depending on the Hamiltonian to be constructed
with real, complex or quaternions numbers).
On the other hand, starting from a semiclassical
analysis, 
Berry suggested that a typical wavefunction
for a chaotic system could be formed by an infinite
superposition of plane waves travelling in random
directions and with random phases\cite{berrypinball}.
Working with this important conjecture,
Berry was able to show that the
probability distribution for those wave functions
is $P(I/<I>)=e^{-I/<I>}$, with $I=\psi \psi^{*}$, 
and that the space-averaged
spatial correlation of the wavefuntion (at a fixed
energy) is proportional to the zeroth order integer Bessel
function. 
Finally, the application of a supersymmetry formalism has
produced a rigorous deduction of the probability
distributions associated to a non-linear supermatrix 
sigma-model\cite{efetov},
that under certain assumptions can be shown to
be equivalent to RMT and results in the
Porter-Thomas distribution\cite{falko}.

It is clear that RMT bears some limitations,
specially derived
from its statistical nature. However, it has the
advantage of providing a
clear object to be studied and a corresponding
well defined methodology. 
In this paper, we adopt this statistical approach,
and having in mind the results from RMT
we study the probability distributions associated
to wave functions relevant for popular surface structure
techniques, like Low-Energy Electron Diffraction (LEED)
or X-ray Photo-Electron Diffraction (PED). 
Wavefunctions have a potentially 
higher information content than the mere inspection
of levels, and they are also the natural objects to be
studied in these techniques.
We find that the computed probability distributions match closely
the statistics of the eigenfunctions of a Hamiltonian
belonging to the Gaussian Unitary Ensemble (GUE).
This is the universality class relevant
to a scattering experiment, i.e., to an open geometry,
where the energy takes values in the continuum
(good quantum numbers characterizing
the wavefunction are the energy and $k_{\parallel}$ to the
surface). 
Therefore, we take a fresh look to the physical system
and advance the hypothesis that the good structural sensitivity
of these techniques can be understood as a manifestation
of quantum chaos on the wavefunctions.
This conjecture is tested for two standard correlation 
factors widely used to measure the {\it distance}
between a reference structure (usually the experimental
one) and a trial one calculated theoretically. 
The results point in the same direction as the statistical
analysis of wavefunctions since we find that there is
a region where the correlation factor grows
at an exponential rate.

Even if our analysis is not directly linked to any
classical analysis, it is worth mentioning that,
the classical analogue of
every quantum system we have considered
behaves chaotically due to the intrinsic
complexity of the many-scatterer
problem\cite{berrypinball,gutzwiller}.
Furthermore, it is interesting to notice that the classical problem
might behave chaotically even if the scatterers are
regularly distributed.
Therefore, it is surprising to find 
such a vast literature on quantum chaos related to
some kind of disorder, but such a little consideration
of ordered systems, because using the classical
analogue to guide intuition, it is not clear
why quantum chaos should not to be found in perfectly ordered
systems. Recently, Mucciolo et al.\cite{mucciolo} have
shown that the high energy region of the calculated
band structure of crystalline Si 
is complex enough to follow the statistical distribution
of levels expected for 
the Gaussian Orthogonal Ensemble (GOE)
universality class.
Inspired by these ideas, we have also presented
a preliminary work studying the statistical properties of 
LEED states\cite{prl} on ordered materials.
Of course, the LEED problem is quite related to
the band structure analysis, 
its main advantage being merely practical because
of the ready availability of experimental data to test
new theoretical findings.


Berry has studied the Sinai's billiard by mapping the problem 
to a periodic array of hard-circles on a plane\cite{berrykkr}. 
This problem can be solved efficiently applying a 
Korringa-Kohn-Rostoker formalism. 
His method not only 
has a number of computational
advantages, but also allows a detailed analysis for 
the different role played by nonisolated and isolated
orbits contributing to the wavefunctions. 
It is interesting to notice that the 
nonisolated orbits add to the
complexity of the system through the boundaries
defining the billiard only. 
Therefore, although their role is non-negligible to determine 
the chaoticity of the levels in the closed system, 
they do not contribute to the open 
problem of scattering. In other words, paths
that never strike a disk do not contribute to
the reflectivity of a surface (like in LEED) or they are
deliberately removed from the analysis
(like in PED or DLEED) 
due to their lack of useful structural information. 
In scattering experiments, these paths
would be characterized by a
probability distribution given by $\chi^{2}_{\infty}$
(a Dirac's delta function).
Therefore, the study of an open system
allows quite naturally to separate the
influence of nonisolated and isolated orbits, because
the nonisolated ones yield only a trivial contribution,
in contradistinction with the essential entanglement
between both types in the bound problem.


The organization of this paper is as follows:
the scattering of a plane wave
by an ordered array of hard-sphere potentials
is analyzed in section II in detail
applying a LEED formalism.
This is a good analogue to Berry's
work on Sinai's billiard from a scattering
point of view, although some important
differences remain (e.g., it is
a genuine three-dimensional system). 
The hard-sphere model is
interesting from a theoretical point of view because of
the strong similarity with the billiard problem, and also
because its analysis uses the same basic tools employed
in the solution of the diffraction by a surface. 
Certainly,
the usual approach to the LEED problem\cite{pendrybook} 
starts computing the diffraction matrices for a single layer, 
and then it proceeds stacking layers by different methods.
In practice, that means solving first the multiple
scattering problem {\it inside} a layer, to solve later on
the multiple scattering problem {\it between} layers
through a stacking process that finally recreates the
material bulk, or at least a thick enough slab.
The use of hard-sphere potentials simplifies
the computational problem allowing the identification of
the key physical elements responsible for the appearance
of the Porter-Thomas probability distribution.
Following the strategy of introducing complexity step by step,
we start computing the reflection and
transmission matrices for one layer of hard-sphere potentials.
Those layers are afterwards stacked to form
an fcc crystal with
an arbitrary lattice parameter borrowed from copper.
Section III gives a similar analysis for
a LEED problem trying to represent realistically
a few selected materials.
Results corresponding to
both Diffuse LEED (DLEED) and
conventional I(E) analysis are discussed in this part.
Finally, we consider in section IV the impact of our
previous findings on the statistical correlation factors
(R-factors) widely used in LEED or PED to assess the confidence on
a structure predicted by theory. This is usually done
by a trial and error fit, comparing theoretically
calculated diffracted intensities with 
the experiment. The usual rules are
as simple as: (i) the lower
the R-factor the better, and (ii) it should represent
a (hopefully global) minimum. In practice, because there
is no way to secure a real global minimum in a multi-dimensional
parameter space, the recipe of 
getting a low-enough value becomes the only guide to trust or not
a given structure. 
We shall see how a new criterion adding to the others
can be obtained
by identifying the existence of a region where
the R-factor changes quickly (exponentially) from
values typically obtained from the application of perturbation
theory ($R \approx 0$)
to values representative for uncorrelated intensities
($R \approx 1$).
We argue that this exponential dependence is a consequence
of the chaotic nature of wave functions obtained in a
complicated multiple-scattering scenario.

\section{Simplified model of electron multiple scattering
by N-scatterer: hard-sphere potentials}

\subsection{Scattering by an isolated potential}

We analyze first the simplest case related to our
problem: the scattering of a plane wave, $e^{i k z}$,
by a single
atomic potential modelled by a hard-sphere of radius
$R$\cite{atomicunits}. 
The scattered wave is given asymptotically by:

\begin{equation}
e^{i k z} + f_{k}(\theta) { e^{i k r} \over r}
\end{equation}

\noindent
with
\begin{equation}
f_{k} (\theta) = {1 \over k} 
\sum_{l=0,\infty} \sqrt{4 \pi (2 l + 1)}
t_{l}(k) Y_{l0}(\theta) 
\end{equation}

\noindent
where,
$$
t_{l}(k) = e^{i \delta_{l}} \sin (\delta_{l}) \; ,
$$
and\cite{mott} 
$$
\delta_{l}(k) = 
\arctan 
\lbrack 
(-1)^{l-1} { 
J_{l+\frac{1}{2}}(k R) 
\over 
J_{-l-\frac{1}{2}}(k R) 
}
\rbrack \; .
$$

\begin{figure}
\epsfig{figure=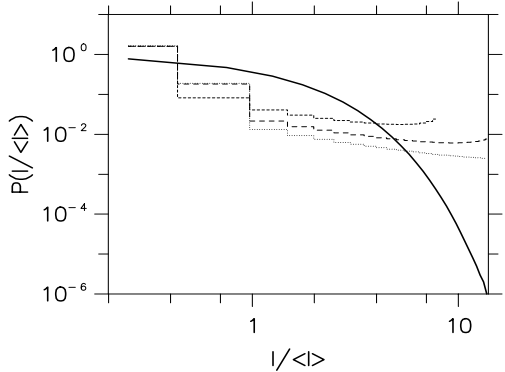,height=10cm,width=8.5cm}
\caption{Probability distribution for the scattering
of a plane wave by a single hard-sphere potential
(R=2 a.u.).
Three energies (in a.u.) are shown: 
$E=2$ (short dashed), $E=5$ (long dashed), and
$E=10$ (dotted). The thick solid line is the $\chi^{2}_{2}$ function
corresponding to the GUE wavefunctions statistics.
}\label{figonev}
\end{figure}

Fig. ~\ref{figonev} displays the probability distribution
function for intensities scattered by this model at
constant energy when the angle $\theta$ is varied.
This is compared with the Porter-Thomas law characteristic
of a chaotic system to stress the different statistical
behaviour. Only two parameters are relevant to this
experiment: the length scale ($R=2$ a.u.),
and the energy scale. The approximate
semiclassical rule $k R \approx l_{max}$
gives us some rough value for the maximum component
in the spherical wave expansion, and twice that value 
is used in all our calculations
($l_{max}=20$). 
Because the phase-shifts bring some
non-trivial dependence on $k R$ 
through the spherical Bessel functions,
results for three different energy
values spanning the range of interest are given
to illustrate this dependence.
Basically, the same statistical pattern is found,
reflecting the smooth variation of the scattering
factor (moduled by the forward peak). 
This probability distribution is remarkably similar
to the one found for the typical wave function of
a chaotic system when too few random components are
used\cite{prl}.

\vskip -5cm

\subsection{Electron diffraction by a plane of
scatterers}

The same statistical probability distribution
is expected for the
diffraction of a single Bravais lattice in the
kinematical approximation at normal incidence,
where all the atoms scatter the plane-wave at the same
time and are equivalent because of the 
Bravais-like symmetry. In this approximation,
the lattice is merely contributing a structure
factor composed of delta functions centered around
Bragg conditions\cite{vanhove2}. 
However, by taking an appropriate limit on
the full dynamical result,
we shall see that this is not the case when the
angle of incidence is varied (cf. Figure ~\ref{figfull_cine}),
because of the $n \times \sin (\theta)$ extra path
added to each scatterer in the plane ($n=0, \infty$).


\vskip -5cm

\begin{figure}
\epsfig{figure=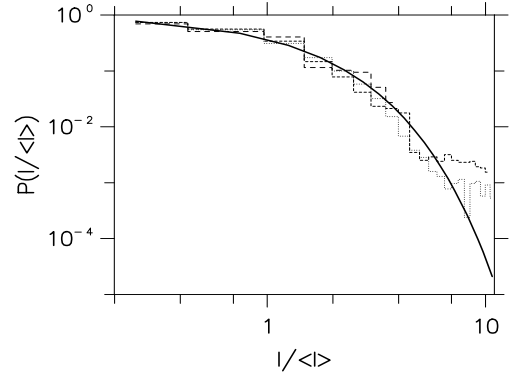,height=10cm,width=8.5cm}
\caption{
Probability distribution of wavefunctions
reflected by one layer of hard-spheres potentials. 
Results for $R=2$ a.u. and three different energies
are shown:
E = $2$ a.u. (long dashed),
$5$ a.u (short dashed), and
$10$ a.u. (dotted).
}\label{figonelay}
\end{figure}

To proceed gradually from simple to more complicated systems,
we now analyze the diffraction matrix of a single 
two-dimensional Bravais
lattice\cite{pendrybook,vanhove2,beeby}:

$$
M_{
\vec K^{\pm}_{\vec g'},
\vec K^{\pm}_{\vec g }} =
{       8 \pi^{2} i \over  
\mid \vec K_{\vec g}^{\pm} \mid
     \vec K_{\vec g',z}^{\pm}   } 
\sum_{lm,l'm'} 
e^{i \delta_{l'}} \sin (\delta_{l'}) \times
$$
\begin{equation}
\{ i^l (-1)^m Y_{l-m}(\vec K^{\pm}_{\vec g })\}
{ 1 \over (1 - X)_{lm,l'm'} }
\{ i^{-l'} Y_{l'm'}(\vec K^{\pm}_{\vec g'})\}
\end{equation}

\noindent
This expression gives the complex amplitude diffracted
from an ingoing beam, $\vec K^{\pm}_{\vec g }$, into
an outgoing one, $\vec K^{\pm}_{\vec g'}$.
Therefore, it is the basic quantity needed to 
compute the reflection, $R=M^{+,-}$ 
and transmission, $T=I+M^{+,+}$,
of just one layer (where $-$ denotes propagation
towards the vacuum where the original wave
was originated, and $+$ in the opposite direction).

\vskip -5cm

\begin{figure}
\epsfig{figure=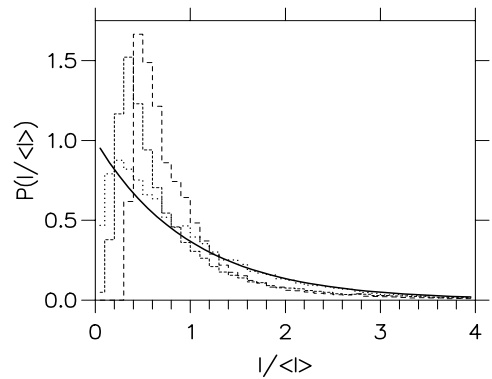,height=10cm,width=8.5cm}
\caption{Dependence of the wavefunction probability distribution
on the hard-sphere radius. Results for
E =10 a.u. and  R=2.0 a.u. (dotted), 0.2 a.u. (dashed)
and, 0.02 a.u. (long-dashed) are shown.
}\label{fig10hvsR}
\end{figure}

Fig. ~\ref{figonelay} shows the wavefunction
statistical distributions obtained for
a two-dimensional square lattice of hard-sphere
potentials at nearest-neighbours distances
taken from a Cu(100) surface ($4.82$ a.u.).
Internal parameters 
relevant for the calculation are
kept to the same values as the preceding case
($R=2$ a.u., and $l_{max}=20$). 
An arbitrarily small positive imaginary part $V_{oi}$ 
is added to the energy to give the Green's functions
the required analytical behaviour. The physical
effect of such a mathematical trick is to ensure
the proper decay of waves at infinity.
The actual value used in our calculations is
$V_{oi}=0.001$ a.u., small enough compared with
all relevant energies as not to
have any noticeable effect
(Kambe's method\cite{kambe} is used
to compute the lattice summation and about 2300
lattice points are included). Given these
values for the internal parameters of the
model,
intensities still depend on external parameters
that typically will be explored
in real experiments: the energy,
and the incident and collection angles.
In our previous work\cite{prl} we have studied
the statistical probability distribution in
scattering intensities varying the collection
angle (Diffuse LEED) and the energy (standard 
I(E) LEED analysis), but at a fixed
incident angle defined by
$\theta$ and $\phi$. 
In this paper, we increase the database size
by considering different initial incident directions
on a solid angle centered around $\theta=0^{\circ}$
and $30^{\circ}-40^{\circ}$ degrees wide.
As the energy is the main external parameter
controlling the experiment in any case,
results for three different energies spanning
the range of interest
($E=2$, $5$ and $10$ a.u.) are 
systematically shown for comparison.
We remark that the trivial limit $\chi^{2}_{\infty}$,
is recovered at fixed energy from Equation (3) for
$l_{max}=0$, or alternatively
when $R \rightarrow 0$, that eventually would make
only one spherical component necessary,
killing all the dependence on the angles $\theta$
or $\phi$. 
None of the parameters used in our model
correspond to unrealistic values regarding
real LEED experiments, except for the very small imaginary
part for the optical potential. In a typical LEED experiment
the inelastic interaction
is strong, concentrating the diffraction process
to the vicinity of the surface (which explains
the sensitivity of the experiment to small
atomic displacements in the few last layers).
This can be taken into account effectively
including a large optical potential 
($\approx 0.1-0.2$ a.u.) in the
energy.
However, it is important to realize that the
experiment is conceived as purely elastic,
and electrons
having lost some energy do not contribute
to the detected intensity.
Realistic values 
have been previously used to make contact with
experiments\cite{prl}, and now we choose to work
in the limit $V_{oi} \rightarrow 0$ to show that
this value is not controlling the resulting
probability distributions in any way.
The actual computer code used is a modern version
of routines given by Pendry\cite{pendrybook}.
Figure ~\ref{figonelay} clearly shows how the
single Bravais layer gives diffracted intensities
that already have statistical distributions
close to the ideal Porter-Thomas
law, irrespectively of the energy.

Next, the energy is fixed at some
arbitrary representative value ($E=10$ a.u.), and the
dependence of the statistical probability distribution
of a two-dimensional Bravais lattice
on the size of the hard-sphere potential is studied.
Fig. ~\ref{fig10hvsR} gives the statistics for three
different sizes of the hard-sphere radius.
As $R \rightarrow 0$, the reflected intensities
also become negligible. From a computational
point of view, we do not expect this limit to be
strictly accessible for a numerical experiment, because
the computer zero, determined by a numerical underflow,
might be contaminated by round-off errors
with a statistical distribution not know {\it a priori}.
Average values for the intensities at $E=10$ a.u. 
are 
$1.6 \times 10^{-2}$, $2.5 \times 10^{-4}$, and
$4.1 \times 10^{-6}$, respectively for 
$R=2$,  $0.2$, and $0.02$ (a.u.).
Figure ~\ref{fig10hvsR} shows the tendency of the
probability distribution towards $\chi^{2}_{\infty}$,
associated with the constant value corresponding to
very small radius value. As the main interest in this case
is to show the behaviour near the origin, we have
skipped the usual log-log plot, well suited to
manifest the fast exponential decay, but not so useful to
stress the behaviour near the origin.

\vskip -5cm

\begin{figure}
\epsfig{figure=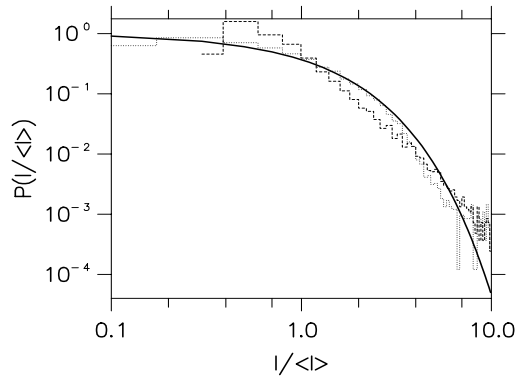,height=10cm,width=8.5cm}
\caption{Full dynamical model for a single Bravais
lattice (dotted) compared to a kinematic one (dashed). 
Parameters for the calculation are: $E=10$ a.u.,
$R=2$ a.u., and $l_{max}= 20$.
}\label{figfull_cine}
\end{figure}

To understand the role of the complexity created by the
intralayer multiple-scattering versus the geometrical
factor of different incident angles,
we have artificially made the
intralayer scattering matrix equal to zero:
$X = 0$.
The result is shown in
Fig. ~\ref{figfull_cine} for fixed values
of $E$, $R$, and $l_{max}$.
It is observed that 
the kinematic N-scatterer problem analyzed for
different incident angles, already approximates 
the Porter-Thomas distribution, 
although the inclusion of intralayer multiple
scattering gives a distribution closer
to the ideal one.

\subsection{Electron diffraction by a stacking of planes of
hard-spheres scatterers}

We use the {\it layer doubling} scheme\cite{pendrybook}
to stack layers of hard-sphere potentials.
Layers are stacked to form an fcc lattice, 
borrowing the intralayer distances
from copper, as before.
Taking into account the small imaginary 
part used in our calculations, we should
at least double the width of the slab up
to distances of about 
$l_{c} \approx {\sqrt{2 E} \over V_{oi} }$,
approximately including 1000 layers.
This is not practical, nor necessary, and we 
only slowly
double the slab size to investigate the influence of
the layer width on the statistical distribution. 

\vskip -5cm

\begin{figure}
\epsfig{figure=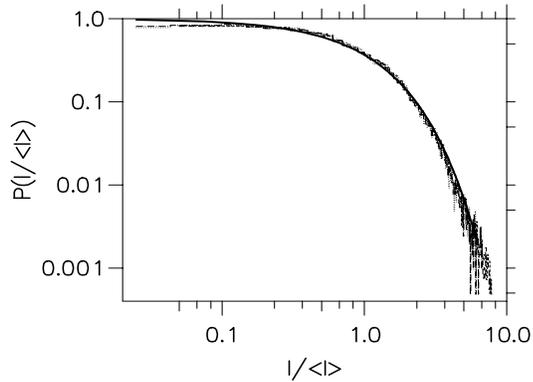,height=10cm,width=8.5cm}
\caption{
Reflected intensities statistics of a full
dynamical calculation of a slab of increasing
width: The number of stacked layers is
$2$ (dotted), $4$ (dashed), $8$ (long-dashed) and, 
$16$ (double-dashed double-dotted).
Other computational parameters are given in the
previous figure caption.
}\label{figvslay}
\end{figure}

This is illustrated in Fig. ~\ref{figvslay}
for four different widths: $2, 4, 8$ and $16$ layers
(for comparison, the result for $1$ layer can be found in
Fig. ~\ref{figfull_cine}).
Other parameters are fixed to 
the same values previously used
to isolate features only associated to the slab width.
It is easily appreciated how the statistical distribution
tends more to behave like the ideal Porter-Thomas one
as the width is increased. It is difficult to
mathematically quantify this tendency, but an approximate way
would be to make a least-squares fit of the data to 
$\chi^{2}_{\nu}$ functions, and give the
$\nu$ with the best agreement.
This results in
$\nu= 2.7, 2.3, 2.2, 2.2$, and $2.2$ 
for $1, 2, 4, 8$ and $16$ layers respectively.
The mean values of the intensity reflected are
for each case: 
$1.58 \times 10^{-2}$,
$2.42 \times 10^{-2}$, 
$2.55 \times 10^{-2}$,
$2.58 \times 10^{-2}$, and 
$2.58 \times 10^{-2}$.
A penetration depth of sixteen layers is to be thought as
a practical upper limit for most experimental systems.
>From these numbers, it must be concluded that,
at least under the particular conditions we have chosen, 
{\it intralayer}
scattering among the first two layers help the
distribution most to compare well with the Porter-Thomas law,
while further stacking of layers mainly contributes to
improve the finer details.
Therefore, a moderate amount of
multiple scattering should be generically
held responsible for the statistical behaviour of wave
functions, characteristic of quantum chaos.





\section{Application to Diffuse and
conventional I(E) LEED analysis}

All these ideas can be used to analyze 
LEED experiments conducted in standard
surface structural analysis. Those experiments
are performed by measuring as a function of the energy,
the available exiting beams
(I(E) standard LEED) or by analyzing many different
exiting beams at a few fixed energies (DLEED).
The former technique is usually 
applied to ordered surfaces, while the latter is
more appropriated to surfaces with disorder.
As the relevant physical principles behind
both techniques are the same, and are well
described by a multiple scattering formalism,
it is not surprising to find that both intensities,
the ones simulated theoretically for realistic
systems, and the corresponding to
experimental measured values,
fit rather well to the Porter-Thomas
distribution. Indeed, the main differences with
our previous model are the atomic potentials
and the important electron-electron
inelastic interaction that attenuates the
wave within a few layers of the surface.
While our results are not very sensitive
to a particular set of phase-shifts, as
become obvious from our results for different
materials simulated with realistic potentials,
the influence of a large optical potential is
balanced by our finding that scattering by one, or two  
layers at most, is enough to reproduce the 
characteristic GUE
probability distribution.

\vskip -5cm

\begin{figure}
\epsfig{figure=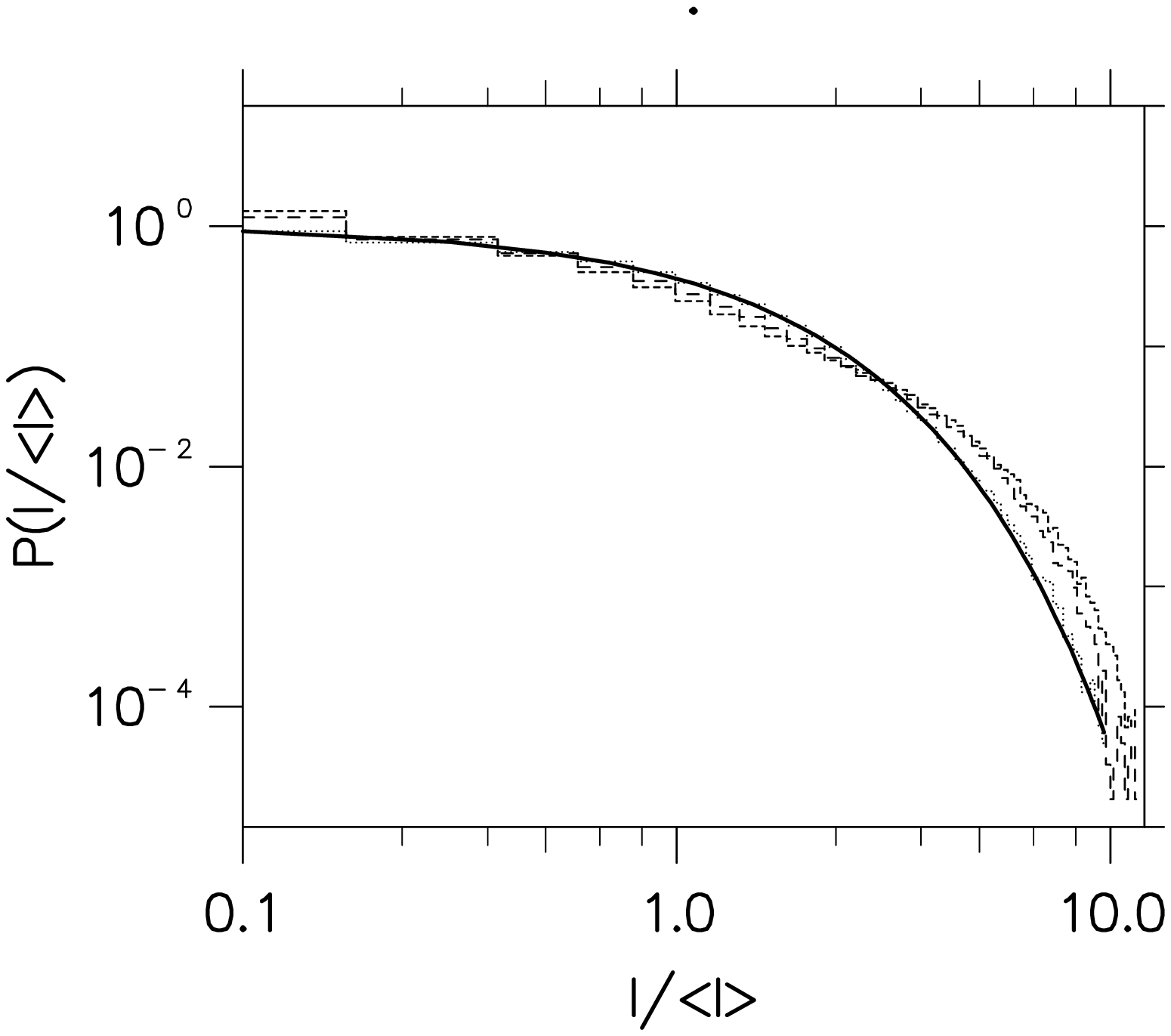,width=8.5cm}
\caption{Probability distribution for DLEED intensities
of O/Ni(100). $E=12$ a.u. (dotted), $14$ a.u. (short dashed),
and $16$ a.u. (long dashed).
}\label{figdleed}
\end{figure}

This is demonstrated in Figures
We first explore the statistical behaviour of
DLEED. A representative system many times
considered in the literature both from
the experimental and the theoretical point of view
is the lattice gas disordered adsorption of
oxygen on Ni(100)\cite{uli}. Nickel is obviously a strong
scatterer and multiple-scattering plays an
important role. This is relevant because
the adsorbate is both illuminated directly by
the wave coming from the electron gun and by the
reflected wave coming from the surface,
that depends on energy and angle in a complicated
manner dictated by the multiple scattering inside
the substrate. Geometrical parameters are taken
from a detailed structural search, fixing the
adsorption position at about 1.5 a.u. on
the four-fold symmetry position in the
square surface lattice (less symmetric
adsorption sites would only make more complex
the system and 
therefore more likely to recover the Porter-Thomas
law). Three different energies 
are computed theoretically using a dynamical
approach at $T=0$ K, as described by
Saldin and Pendry\cite{cpcdleed}.
To improve the statistics, different incident angles
$\theta$ and $\phi$ are computed and used 
as explained above.
The agreement with Porter-Thomas is quite good 
(e.g., $E=12$ a.u.
gives a least-squares fit $\nu$ value of $2.0$),
reflecting that although the diffuse
background is generated by scattering with only one atom
responsible for breaking the otherwise perfect symmetry 
that would result in
Bragg conditions, the wave reflected by the substrate
and illuminating the atom 
is very complicated. 
Obviously that what matters most here is
that the forward scattering of that wave by the adsorbate is
stronger or comparable to the backscattering of the simple
plane wave by the same potential. 

\begin{figure}
\epsfig{figure=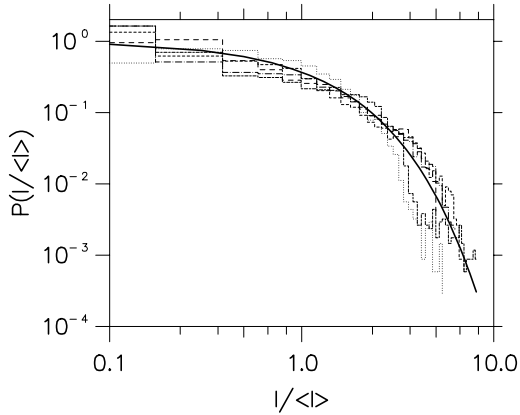,width=8.5cm}
\caption{Probability distribution for 
experimental DLEED intensities
of O/Ni(100). 
$E=3.7$ a.u. (dotted), 
$5.6$ a.u. (short dashed), 
$7.5$ a.u. (long dashed),
$9.4$ a.u. (long dashed-dotted), and
$11.1$ a.u. (long dashed-short dashed).
}\label{figdleedo}
\end{figure}

Our previous findings can also be corroborated analyzing
experimental DLEED intensities measured for the same
system by the Erlangen group\cite{klausdata}.
Figure ~\ref{figdleedo} shows such an analysis for
five different energies going from $E=3.7$ a.u. to
$E=11.1$ a.u. in approximate steps of $1.8$ a.u.
The database was measured
in a sample cooled down to 
liquid Nitrogen temperature ($\approx 90$ K),
at normal incidence, and for an approximate
coverage of $0.25$ ($1$ representing one 
full monolayer adsorbed).
Less than one thirtieth 
of the amount of data used
in the theoretical analysis is available, making
a poorer statistics, but the expected tendency
is followed well, although fluctuations are 
clearly observed.

\vskip -5cm

\begin{figure}
\epsfig{figure=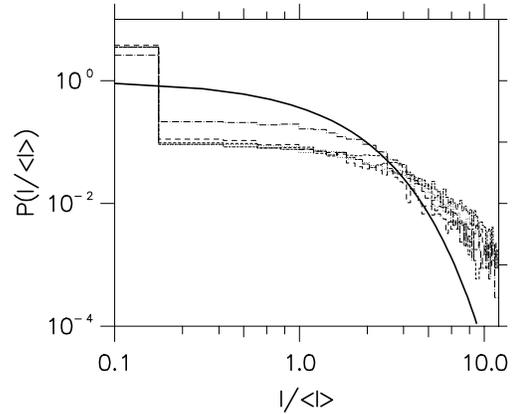,width=8.5cm}
\caption{Probability distribution for 
experimental DLEED intensities
of K/Ni(100). 
$E=3.7$ a.u. (dotted), 
$5.6$ a.u. (short dashed), 
$7.5$ a.u. (long dashed),
$9.4$ a.u. (long dashed-dotted), and
$11.1$ a.u. (long dashed-short dashed).
}\label{figdleedk}
\end{figure}

We have also analyzed the experimental data
measured by the Erlangen group on
the system: K/Ni(100) -disordered-.
Normal incidence and liquid Nitrogen cooling ($T=90$ K)
are used again. Potassium coverage is kept
at a lower value of $0.05$. 
This system, however, shows an important
difference with the last one: Potassium is
adsorbed at the hollow site, but at a
much higher position, $\approx 5.1$ a.u.
It is clear that the source of complexity is
the substrate, and if the atom would be isolated,
or too far away from the surface, the DLEED
intensities would simply correspond to the ones
due to a single atomic potential.
>From  our previous estimate for the
typical decaying length, we find that
$l_{c} \approx 15-20$ a.u. for the energies
involved in the experimental data. Therefore,
it is not unexpected that the probability distribution
for K/Ni(100) bear some similarity to the one
obtained for an isolated atom, as can be seen
in Figure ~\ref{figdleedk}.

\vskip -5cm

\begin{figure}
\epsfig{figure=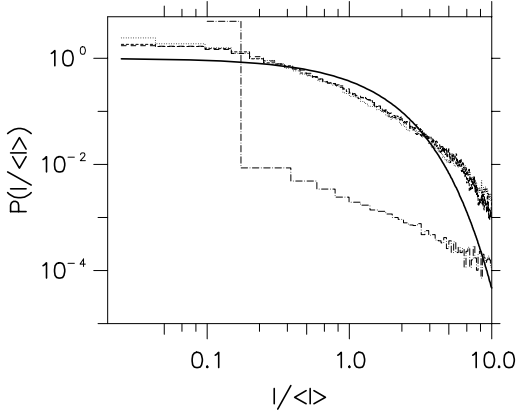,width=8.5cm}
\caption{Probability distribution corresponding
to the LEED I(E) curves for Cu(100).  considering 
different surface relaxations, $d_{12}=3.40$ a.u. (dotted),
$d_{12}=3.02$ a.u. (short dashed), an equally weighted average 
from $3.02$ to $3.74$ (long dashed), 
and $d_{12}=2.00$ a.u. (dotted-dashed).
}\label{figleedivcu}
\end{figure}

\begin{figure}
\epsfig{figure=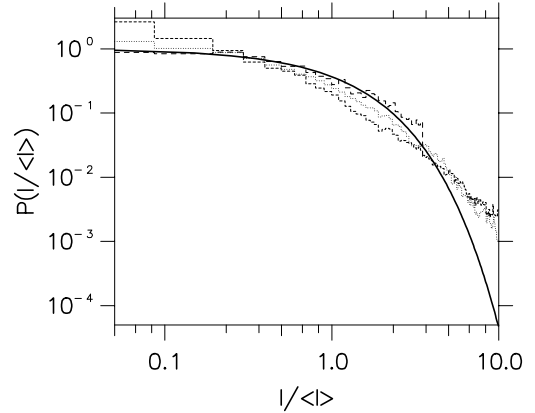,width=8.5cm}
\caption{Probability distribution corresponding
to the LEED I(E) curves for W(100) (dotted), 
Si(111) (dashed)
and c($8 \times 2$)-GaAs(100) (long dashed).
}\label{figleedivgaas}
\end{figure}

By reference to the RMT, or to Berry's hypothesis
about the structure of the typical wavefunction
on a chaotic system, it is clear that the
same Porter-Thomas law should manifest if
intensities are analyzed as a function of
energy at a fixed arbitrary position, $\vec r$.
We have analyzed this behaviour for scattering
wavefunctions by calculating 
LEED I(E) curves for three materials not bearing 
a structural, nor an electronic relationship:
Cu(100), W(100) and Si(111), and by analyzing
a real LEED experiment. 
Figure ~\ref{figleedivcu} shows the
probability distributions computed for Cu(100),
and Figure ~\ref{figleedivgaas} gives the same
result for W(100), Si(111) and experimental
data for c($8 \times 2$) GaAs(100).
The Van Hove-Tong LEED package is used
to compute intensities within
the Renormalized Forward Scattering (RFS)
approximation to describe multiple scattering
between layers,
and realistic
phase-shifts representing the atomic potentials
are considered\cite{vhthin}.
To improve the statistical confidence of the
results, we chose an arbitrary azimuthal angle
(not related to any symmetry direction)
of $\phi=30^{\circ}$, and we explore a range
of polar angles
from $\theta=5^{\circ}$ to 
$\theta=40^{\circ}$,
in steps of  
$\theta=5^{\circ}$. 
The first $9$ emergent beams
over an energy range from $E=2$ to $20$ a.u. are
considered and
other parameters relevant for the calculation are:
$l_{max}=7$, $V_{oi}= .15$ a.u.,
$T=0$ K, and up to a maximum of
101 beams included.
For copper, we consider
two different distances between the first and
the second surface layers, $d_{12}=3.4$ a.u. and
$3.02$, corresponding respectively to the perfect unrelaxed
surface and to the experimental relaxation
found on clean Cu(100) crystals.
Both cases are seen (cf. Figure ~\ref{figleedivcu})
to be well represented by the $\chi^{2}_{2}$
probability distribution.
An average of intensities for 
different samples with $d_{12}$ values going
from $3.02$ to $3.74$ in steps of $0.08$ a.u.
is also considered with a similar result.
In addition, we compute for an hypothetical relaxation of
$d_{12}=2.0$ a.u., where the RFS 
technique is used outside its validity
region and it results in unphysical divergences.
The statistical distribution associated with
this absurd case is seen to be very different
from the previous ones, signaling clearly that
something went wrong in the calculation.
This is a extreme situation, 
but the same we have been able to
detect a theoretical problem with a simple
statistical analysis, the Porter-Thomas
distribution may help to identify
cases where gross experimental systematic
errors, like an improper subtraction of the
background, saturation of some bright beams,
etc, occur.
We have repeated 
a similar theoretical analysis for W(100) and
Si(111) surfaces in Figure ~\ref{figleedivgaas}.
While similar conditions are used for W(100),
owing to experimental practical difficulties to
measure the high energy end in semiconductors,
we use a smaller energy range for Si(111)
(from $1.$ to $11$ a.u.), but the first $13$ emerging
beams are considered to have a database with 
a similar size to the one considered for the metals.

\vskip -5cm

Although agreement between the multiple-scattering
calculations and experimental data is very good
for geometries representing well a given 
surface (e.g., Pendry R-factor for 
the structural analysis of Cu(100) is
already below 0.1, and experimental and theoretical
I(E) curves are hardly distinguishable by simple
ocular inspection), we also take into account experimental
intensities from 
the c($8 \times 2$)-GaAs(100) reconstruction\cite{palomares}.
Nineteen independent beams measured at
normal incidence, and giving an approximated
energy range of $86$ a.u. are considered.
Intensities have been digitalized from the published results,
and interpolated with splines 
in such a way that the interpolated curves cannot be distinguished
by eye from the measured data.
Again, a fair agreement between the obtained
probability distributions and the ideal Porter-Thomas
law is obtained, although we observe that the 
agreement is achieved over a limited range
due to the smaller amount of available data.

\vskip -5cm

\section{R-factors dependence on geometrical
structural parameters}

An interesting question derived from the 
chaotic nature of the wave function is how quickly 
two given structures differing in a given
structural parameter, $p$, become unrelated from
the point of view of diffracted intensities.
This process happens, although
the two structures are always related through some underlaying
geometrical relationship, via the complexity introduced by multiple
scattering.
To demonstrate this point, we first analyze
theoretically computed diffracted intensities
where only one relevant parameter is varied.
Afterwards, we apply the same ideas to a recent 
structural search on an experimental system performed
by other people\cite{polop}. The theoretical experiment
is performed using the same DLEED program
mentioned before\cite{cpcdleed}. 
This allows us to simulate the adsorption of an
oxygen atom on the hollow site of a perfect
(unrelaxed) Ni(100)
surface. The reference height is fixed again at 1.5 a.u. 
from the layer defined by the
nickel cores, which is very similar to the
experimental value. Changing the oxygen adsorption height
we study the corresponding changes in the DLEED
intensities.
We notice that the same code has been previously used
for a real structural search on this system,
proving its capability to give a realistic description
of the physical system\cite{uli}.

To measure the changes in the LEED diffracted
intensities we adopt two common but otherwise
unrelated correlation factors:
(i) the Root Mean Squared Deviation
$R_{RMSD}$\cite{vanhove2}, and (ii) the
Pendry R-factor, $R_{P}$\cite{pendryrp}. 
The former is the simplest choice at hand,
and we apply it to the DLEED theoretical experiment:

\begin{equation}
R_{RMSD} = \sqrt{ {1 \over N} \sum_{k=1,N} (I^{ref}_{k}-I_{k})^2 }
\end{equation}

\noindent
where $k$ labels the different $\vec k_{\parallel,out}$.
This R-factor is conveniently normalized 
to ${1 \over \sqrt{N}}$,
the value expected for two random sets of intensities
with the same average value (intensities are 
normalized to their average value).
On the other hand, the Pendry R-factor is a very 
common choice in standard
structural analysis of I(E) LEED curves,
and because it was used by 
Polop et al.\cite{polop} in their study of the
c(2 $\times$ 2)-Si/Cu(110), we simply analyze the
behaviour of their published values. The fact that
we find the same type of behaviour with two so different
R-factors supports our hypothesis that the effects
discussed in this section are quite general. 

\vskip -5cm

\begin{figure}
\epsfig{figure=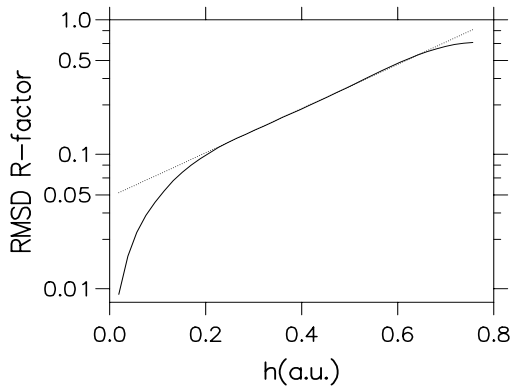,height=10cm,width=8.5cm}
\caption{
Root mean square deviation ($R_{2}$)
measuring the correlation between
a reference structure, $S(0)$, and
a family of structures labelled by
the adsorption height of an atom,
$S(h)$.
Region II (see text) is shown as the linear region
(in logarithmic scale) between the minimum (region I)
and the saturation region (region III).
Energy is fixed to $2.0$ a.u.
}\label{figrms2}
\end{figure}

\begin{figure}
\epsfig{figure=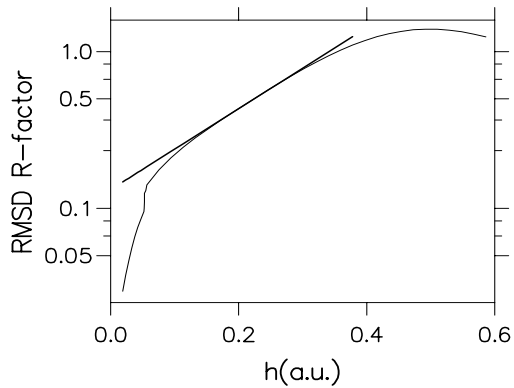,height=10cm,width=8.5cm}
\caption{
Same results as presented in
}\label{figrms5}
\end{figure}

In a previous publication\cite{prl}, we have
distinguished the existence   
of three different regions in parameter
space, $P$: 
(I) a perturbative region,
characterized by a polynomial dependence,
$R(p) \propto p^{n}$; 
(II) an exponential region,
$R( p) \propto e^{p}$,
where small changes in a given structural
parameter result in rapidly increasing
R-values;
and (III) a fully chaotic region where the
R-factor saturates approximately to the values expected
for the comparison between two randomly
generated structures (by
definition $\approx 1$ in both R-factors used here).
Beyond these regions,
the existence of multiple coincidence minima
recreates again similar, although more imperfect
conditions\cite{andersson}.
Existence of region I is justified by
the applicability of perturbative techniques
like Tensor LEED. 
The range of validity of the simplest version of
perturbation theory (e.g., Tensor LEED in its first
form, where the perturbing potential
upon displacement of one atom is simply proportional
to that displacement) is known by common experience
on different systems to be $\approx 0.2$ a.u.\cite{rousphd}.
It is interesting to notice that more sophisticated versions
of TLEED could extend their range of validity to
$\approx 0.4-0.6$ a.u., being 
practically impossible to extend the use of perturbation
theory beyond $0.8$ a.u. We remark that those
displacements correspond typically to R-factors 
between the reference structure (by definition $R=0$)
and the one to be computed perturbatively
of about $R=0.2-0.4$.
In the examples presented below 
(figures ~\ref{figrms2} and ~\ref{figrms5}),
we observe that this normally corresponds to a
region where the R-factor change exponentially with
the structural parameter. Therefore, 
it is possible to understand
why perturbation theory completely breaks 
within this region in spite of
the considerable effort 
that has been made to apply it\cite{cartesian}. 

\vskip -5cm

\begin{figure}
\epsfig{figure=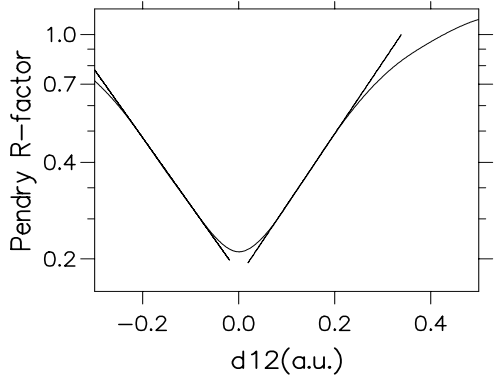,height=10cm,width=8.5cm}
\caption{
The Pendry R-factor is analyzed in a similar way to $R_{RMSD}$ 
in Figure ~\ref{figrms2}. Data corresponding to an experimental 
structural analysis by I(E) LEED of 
${\rm c}(2 \times 2) {\rm Si/Cu(110)}$ is considered.
}\label{figrp}
\end{figure}

The existence of region II
is illustrated
in figures ~\ref{figrms2}, and ~\ref{figrms5}
with theoretically simulated data,
and in ~\ref{figrp} with data produced by
comparing with the experiments.
Regarding the theoretical simulation,
we have slowly increased the
adsorption height of the oxygen atom to
get a corresponding increasing in the $R_{2}$.
To show the functional dependence between
$R_{2}$ and $h$, we take logarithms in
the ordinate axis and we identify a region
where the curve can be approximated very well
by a straight line.
This region 
should be considered the onset of quantum chaos,
and therefore a region of dubious value for
structural work.

\vskip -5cm

To test whether this behaviour is particular to
a given definition of the R-factor or not, we perform the
same analysis using the Pendry R-factor. This
function is defined in very different terms to the
simpler root mean squared $R_{2}$ considered above.
However,
Fig. ~\ref{figrp} shows the same type of behaviour 
for $R_{P}$.
This corresponds to R-factors comparing structural
models to experimental data for
a recent structural search performed on the
system 
${\rm c}(2 \times 2) {\rm Si/Cu(110)}$ 
using conventional I(E) curves\cite{polop}.
$R_{P}$ is studied
as a function of the two
outer inter-layer distance, $d_{12}$,
where the best value provided by the structural
work has been subtracted to put the origin
at $d_{12}=0$.

\section{conclusions}

Scattering intensities have been analyzed from a
statistical point of view. The computed
probability distributions compare well with
the Porter-Thomas law, typical of
random wavefunctions. To understand the origin
of such a similarity we have analyzed models
with increasing scattering complexity, using
a hard-sphere approximation for the interaction
potentials.
The simplest case found displaying a statistical
distribution similar to the
Porter-Thomas law is single scattering by 
a Bravais lattice of N-scatterers 
at an arbitrary angle of incidence.
When more complexity is added to
the system (e.g., considering intra-layer multiple scattering,
or multiple scattering between a
few layers through a {\it layer
doubling} stacking strategy),
the statistical distribution
of such idealized systems shows
better agreement with the ideal $\chi^{2}_{2}$
function.
The same behaviour is found if realistic
potentials are considered to describe the atoms within
the periodic lattice. 
The analysis of real experimental data is also
consistent with the same ideas, as expected from
the known reliability of those theoretical methods
to give scattering intensities if the geometries are
already known.
Finally, we have found that standard R-factors behave
exponentially in a transition region (II), before the
intrinsic complexity of multiple scattering effectively
decouples wavefunctions for different geometrical
structures. The existence of this region allows us
to use concepts borrowed from classical chaos,
and to propose a new criterion
for the reliability of a given minimum in the R-factor,
depending on whether the structure lies in
a perturbative region (I), or beyond the transition
zone (III).

\section{acknowledgments}
This work has been supported by the CICYT under
contracts num. PB96-0085 and PB97-1224.
We are grateful with Prof. K. Heinz for
making available to us the experimental DLEED
data.


\begin{thebibliography}{10}
\bibitem{atomicunits}
$\hbar=m_{e}=e=1$ is
used 
throughout the paper
(i.e., energy is expressed in Hartrees, 27.2 eV,
and distances in Bohrs, $0.0529$ nm).
\end{thebibliography}
\end{document}